\documentclass[aps,pre,10pt,twocolumn]{revtex4-1}
\usepackage[font=small,labelfont=bf]{caption}
\usepackage{amsmath}
\usepackage{amsfonts}
\usepackage{amssymb}
\usepackage{graphicx}
\usepackage{natbib}
\newcommand{\be}{\begin{equation}}
\newcommand{\bea}{\begin{eqnarray}}
\newcommand{\bc}{\begin{center}}            
\newcommand{\ee}{\end{equation}}
\newcommand{\eea}{\end{eqnarray}}
\newcommand{\ec}{\end{center}}

\newcommand{\baa}{\begin{eqnarray*}}
\newcommand{\eaa}{\end{eqnarray*}}
\begin{document}
\title{Thermoelectric engines with internal and external irreversibilities
at optimal power}
\title{Optimal power analysis of thermoelectric generator with rectified Joule 
heating
and one-sided external irreversibility}
\title{Thermoelectric generator at optimal power 
with external and internal irreversibilities }

\author{Jasleen Kaur}
\email{jasleenkaur@iisermohali.ac.in}
\author{Ramandeep S. Johal}
\email{rsjohal@iisermohali.ac.in}
\affiliation{ Department of Physical Sciences, \\ 
Indian Institute of Science Education and Research Mohali,
Sector 81, S.A.S. Nagar, Manauli PO 140306, Punjab, India}
\begin{abstract} 
There are few exact results on optimal power conditions for a thermoelectric
 generator in the presence of both external and internal 
 irreversibilities---modelled as non-ideal thermal 
 contacts and Joule heating, respectively.
 Simplified cases, where only one kind  of irreversibility is assumed, yield
some well-known expressions for efficiency at maximum power (EMP), such as 
Curzon-Ahlborn
 efficiency for endoreversible model. In this work, we analyze situations under
  the simultaneous presence of internal and external irreversibilities. To 
simplify,
   we neglect heat leaks, and each kind of irreversibility is assumed only 
   on the side of one of the thermal
  contacts. We also present the symmetric case---where each kind of 
irreversibility
   contributes with equal strengths towards the side of each thermal contact. We 
show the bounds satisfied by EMP in each of these
 regimes and compare its properties for thermal impedence matching and
  close to equilibrium, where we find step-wise changes in EMP. 
\end{abstract} 
\maketitle
\section{Introduction}
Sadi Carnot proposed an ideal heat engine exploiting the temperature difference
 between two heat reserviors \cite{Carnot1960reflections}. However, the presence
  of various irreversibilities makes it impractical to design such an engine.
A thermoelectric generator (TEG) provides a paradigmatic model of a realistic heat
 engine, where both internal  and external sources of irreversibilities
 can be considered \cite{Gordon1991, Apertet2012B}. Along with a sustained effort
 towards improving the figure of merit of a thermoelectric material (TEM) 
 \cite{Nemir2010, Littman1961, snyder2011,majumdar2004,shakouri2011}, the
  characterization of optimal performance also forms a significant aspect of the
 study of thermoelectricity 
 \cite{RIFFAT2003,DiSalvo1999,Pei2011,Goupil2011,Goldsmid2010,Rowe1995,Snyder2003}.
 The framework of finite-time thermodynamics aims to characterize the
 performance of thermal machines with finite-rate processes 
 \cite{Berry1984,DavidJou2008,Andresen2011,Salamon2001,BENENTI2017}.
 
 An important quantity in this regard is the efficiency at maximum power 
 \cite{VandenBroeck2005,Esposito2009,Ouerdane2015,Moreau2012,Wangtu2013}, 
 referred to as EMP. A simplified analysis may be done by considering either external or
   internal irreversibility \cite{Agrawal1997,Apertet2013A}.
   Further, heat leaks can be neglected by using the strong-coupling assumption 
   \cite{Esposito2009, Apertet2012B}. In the so-called endoreversible 
    model \cite{Chambadal1957,Novikov1958}, 
    an external irreversibility is caused by a finite rate of heat transfer
  between the working substance (TEM) and heat 
reservoir. Based on Newtonian heat flows, this model yields EMP 
 \be
 \eta_{\rm ext} = 1-\sqrt{1-\eta_{C}^{}},
 \ee 
 which was introduced in physics literature by Curzon and Ahlborn \cite{CA1975, Agrawal1997}.
 Here, $\eta_{C}^{} = 1-T_{c}^{}/T_{h}^{}$ is Carnot efficiency, with $T_c$ 
$(T_h)$ is the temperature of cold (hot) reservoir. This value is independent of any 
properties of TEM, or of thermal contacts.
  
 On the other hand, thermal contacts between the working substance and a
 heat reservoir may be perfect, while Joule heating within TEM acts as the 
source of internal irreversibility. A different expression for EMP is then obtained
 \be
 \eta_{\rm int}^{}=\frac{\eta_{C}^{}}{2-(1-\omega) \eta_{C}^{}},
 \label{SS}
 \ee
 which can be obtained in other models too \cite{SchmiedlSeifert2008,Chen1989,
Johal2018}. Here, parameter $\omega$ is the fraction of Joule heat rejected to the cold 
reservoir see Eq. (\ref{cflux}). For a homogeneous TEM, a value of $\omega =1/2$ is expected 
\cite{Rowe1995}. 
 
Apart from the above idealized cases, the exact analysis of optimal performance
is not straightforward in the general scenario \cite{Gordon1991, {Apertet2012A}}. In this 
work, we highlight a few special cases which still lead to a tractable problem in power 
optimization. Thus, the external irreversibility may be considered only at one thermal 
contact while the other contact is assumed ideal \cite{Novikov1958}. This requires 
tuning the thermal
  conductances of contacts with reservoirs. Secondly, we assume  that Joule heat
  is fully transferred to one of the heat reservoirs \cite{Tingyu2014,Chang2006}, i.e. 
$\omega=1$, or $0$. With advances in the
   fabrication of functionally graded thermoelectric materials 
 \cite{Rowe1995,Ioffe1957semiconductor}, it is possible to rectify
  Joule heat such that the proportion of Joule heat flowing into a reservoir can
   be controlled. These assumptions allow for an exact analysis of  optimal 
power. In addition to that, we discuss the ``symmetric'' case, where internal and
     external irreversibilities contribute equally  on both sides of TEM.
     For all these cases, we obtain exact 
expressions for EMP which
      depend on the ratio of cold to hot temperatures, as well as on the ratio
       of the external to internal thermal conductances. Further, EMP is limited
        within certain bounds that may be approached by suitably tuning the 
ratio of thermal conductances.

Our article is organized as follows. In Section II, we describe the model of a
 TEG. In sections III, IV, V and VI, we discuss TEG having different 
combinations of external and internal irreversibilities and optimize the power output. In
   Section VII, we discuss the TEG having symmetric contributions of internal 
and external dissipation. We end with a discussion of results in Section VIII 
    and concluding remarks in Section IX.
\section{TEG model}
Thermoelectricity is a non-equilibrium phenomenon, which can be studied within
 the framework of Onsager-Callen theory 
 \cite{Onsager1931,Callen1948}. The coupling
  between the gradients of temperature and electric potential gives rise to
   various thermoelectric effects \cite{pottier2014, Callenbook1985}.
 We consider TEM to be a one-dimensional substance with given values of
 internal resistance $R$ and Seebeck coefficienct $\alpha$. Further,
 let $I$ denote the constant current flowing through the TEM 
 (see Fig. 1). Then, based on Onsager formalism and
   Domenicali's heat equation \cite{Domenicali1954, Apertet2013B}, thermal fluxes at 
the end points of TEM are written as follows.
 \bea
 \dot{Q}_h^{} &=& \alpha T_{hM} I +K (T_{h}-T_{c})-(1-\omega)RI^{2}, 
\label{hflux}\\
 \dot{Q}_c^{} &=& \alpha T_{cM} I +K(T_{h}-T_{c}) + \omega RI^{2} \label{cflux}.
 \eea
 \begin{figure}
	\includegraphics[width=9cm]{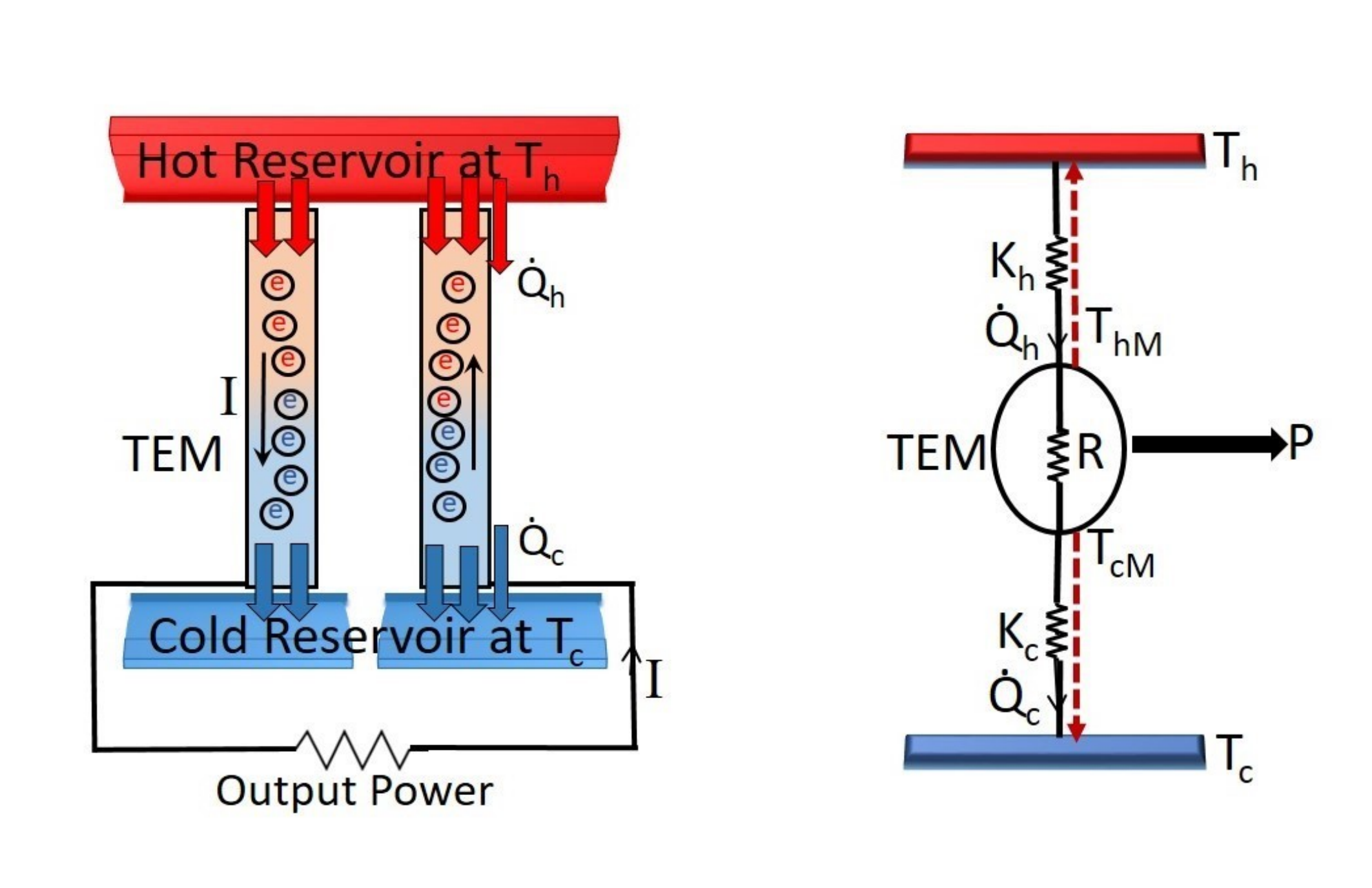} 
	\caption{ A TEG consists of two legs of TEM which are 
	connected electrically in series and thermally in parallel.  
On right side, is a block diagram of TEG having external thermal conductances $K_h$ and 
$K_c$.  $R$ is the internal resistance of TEM with electric current $I$ flowing through.
	$T_{hM}$ and $T_{cM}$ are the local temperatures of TEM towards 
	the hot and cold side respectively. 
	Dashed lines indicate flow of Joule heat into each reservoir.}
	\label{fig1}
\end{figure}
In the above equations, the first term corresponds to 
 convective heat flow, where $T_{hM}$ $(T_{cM})$ is the local temperature of TEM
  at hot (cold) side. The second term signifies heat leakage between the 
reservoirs, 
    and the last term is the fraction of Joule heat received by each reservoir.
     Out of these, the second term has no contribution towards energy conversion 
      \cite{VandenBroeck2005}, and so we work within the strong-coupling
       assumption i.e., 
       $K=0$ \cite{Apertet2013A}. Further, we assume a Newtonian heat flow between 
a reservoir and TEM, whereby we have
\bea
\dot{Q}_h &=& K_{h}^{}(T_{h}^{}-T_{hM}^{}), 
\label{qhnew}
\\
\dot{Q}_{c} & =& K_{c}^{}(T_{cM}^{}-T_{c}^{}).
\label{qcnew}
\eea
  Then, the flux-matching condition on the hot side of TEM gives
\begin{equation}
K_{h}^{}(T_{h}^{}-T_{hM}^{})=\alpha T_{hM}^{}I -(1-\omega)R I^{2} \label{eq5}.
\end{equation}
On solving for $T_{hM}$, and substituting in Eq. (\ref{qhnew}), we obtain
\begin{equation}
\dot{Q}_{h}^{}=K_{h}^{} \frac{\alpha T_{h}^{}I-(1-\omega)R
I^{2}}{K_{h}^{}+\alpha I}. 
\label{eq6}
\end{equation}\\
Similarly, flux-matching condition on the cold side is given as:
\begin{equation}
K_{c}^{}(T_{cM}^{}-T_{c}^{})=\alpha T_{cM}^{}I+\omega RI^{2}.
\end{equation}
So, the outgoing flux is:
\begin{equation}
\dot{Q}_{c}^{}= K_{c}^{} \frac{\alpha T_{c}^{}I+\omega 
	R I^{2}}{K_{c}^{}-\alpha I}. 
\end{equation}
Now, the power output of the device is given by:
\be
P =\dot{Q}_{h}^{}-\dot{Q}_{c}^{},
\ee
with the efficiency $\eta = P / \dot{Q}_{h}^{}$. It
is cumbersome to optimize power of the TEG having both internal and external
irreversibilities \cite{Apertet2012A} i.e., with finite
values of both $K_h$ and $K_c$, as well as for general values of $\omega$. In
 this paper, we treat a few exactly solvable cases, which are interesting in 
their own right:
  
Case 1. Finite value of thermal conductance $K_h$, while the cold contact is
 reversible ($K_c \to \infty$). Additionally, $\omega=0$, implying that the 
Joule heat is totally transferred to the hot reservoir (see Fig. 2a).
  
Case 2. Finite thermal conductance $K_c$, while hot contact as reversible
 ($K_h \to \infty$). Additionally, $\omega=1$, implying all the Joule heat is
 transferred to the cold reservoir (see Fig. 2b). 

Case 3. Finite thermal conductance $K_h$, along with  $\omega=1$, implying
all the Joule heat is transferrred towards the cold reservoir (see Fig. 3a).

Case 4. Finite thermal conductance $K_c$. Additionally, $\omega=0$, implying 
that all the Joule heat is transferred towards the hot reservoir (see Fig. 3b).
 
Case 5. Symmetric external and internal irreversibilities, implying $K_h =K_c$
 and $\omega =1/2$.

We show that in all the above cases, we can analytically optimize power
 with respect to current $I$ flowing through TEM. It is convenient to
  define a parameter $v$ as the ratio of external to internal thermal 
conductances:
\be
v = \frac{K_{\rm ext}}{K_{\rm int}},
\label{defv}
\ee
where, we have \cite{Apertet2012C} 
\bea 
K_{\rm ext} &=& \frac{K_h K_c}{K_h + K_c}, \\  
K_{\rm int} &=&  \frac{\alpha^2}{R} \left[ \omega T_h + (1-\omega)T_c\right].
\eea
In the following, we are able to find compact expressions for EMP, which are functions
 only of $\theta \equiv T_c/T_h$ and parameter $v$ for each respective case. 
\section{Case 1: Finite $K_h$ with $\omega=0$} \label{section3}
In this case, the internal and external irreversibilities
are considered only on the hot side, while irreversibilities on the cold side 
are regarded null. Thus, $\omega =0$, and Eq. (\ref{eq5}) yields  
\be
 T_{hM}=\frac{K_{h}T_{h}+R I^2}{K_{h}+\alpha I}, 
\ee so that 
\begin{eqnarray}
\dot{Q}_{h}^{}=\frac{\alpha T_{h}^{}K_{h}^{}I-R K_{h}^{} I^{2}}{K_{h}^{}+\alpha I}.
\label{LT1}
\end{eqnarray}
Also, thermal flux on the  (reversible) cold side  is 
\begin{equation}
\dot{Q}_{c}^{}=\alpha T_{c}^{} I \label{LT2}.
\end{equation}
Optimizing $P$ w.r.t $I$, i.e. setting $\partial P/ \partial I = 0$, we  solve
 the resulting extremum condition to obtain EMP as 
\begin{equation} \label{eq.12}
\eta^{*}=\frac {v + 1-\sqrt{\theta(v+1) (v\theta+1)}}{v +v \theta+1}, 
\end{equation}
where the parameter $v$ (Eq. (\ref{defv})) in this case is
\be
v = \frac{R K_h}{\alpha^2 T_c},
\label{defv1}
\ee
with $K_{\rm ext}=K_{h}$ and $K_{\rm int}= {\alpha^2 T_{c}^{}}/{R}$. 

Now, we may analyze the behaviour of EMP as a function of $v$. For a given 
$\theta$,
 ${\eta}^*$ is a monotonic increasing function of $v$. We consider coefficient
  $\alpha$ to be fixed. Consider the regime $K_{\rm ext} \ll K_{\rm int}$,
  which implies that the external irreversibility is dominant over the internal
  irreversibility. In other words, we have the limit
    $v\to 0$, or, operationally it implies $R\to 0$. 
    Note that  if $v \to 0$ implies $K_h\to 0$,
     then it yields a vanishing power, and we may exclude this possibility.
 Thus, with a finite $K_h$, the limit $v \to 0$ implies zero Joule heating, and
  so we approach the endoreversible model---an external
   irreversibility on the hot side only. The corresponding EMP is:  $\eta^{*}=\eta_{\rm ext}$.
\begin{figure}
\centering
\includegraphics[width=8.5cm, height=6.5cm]{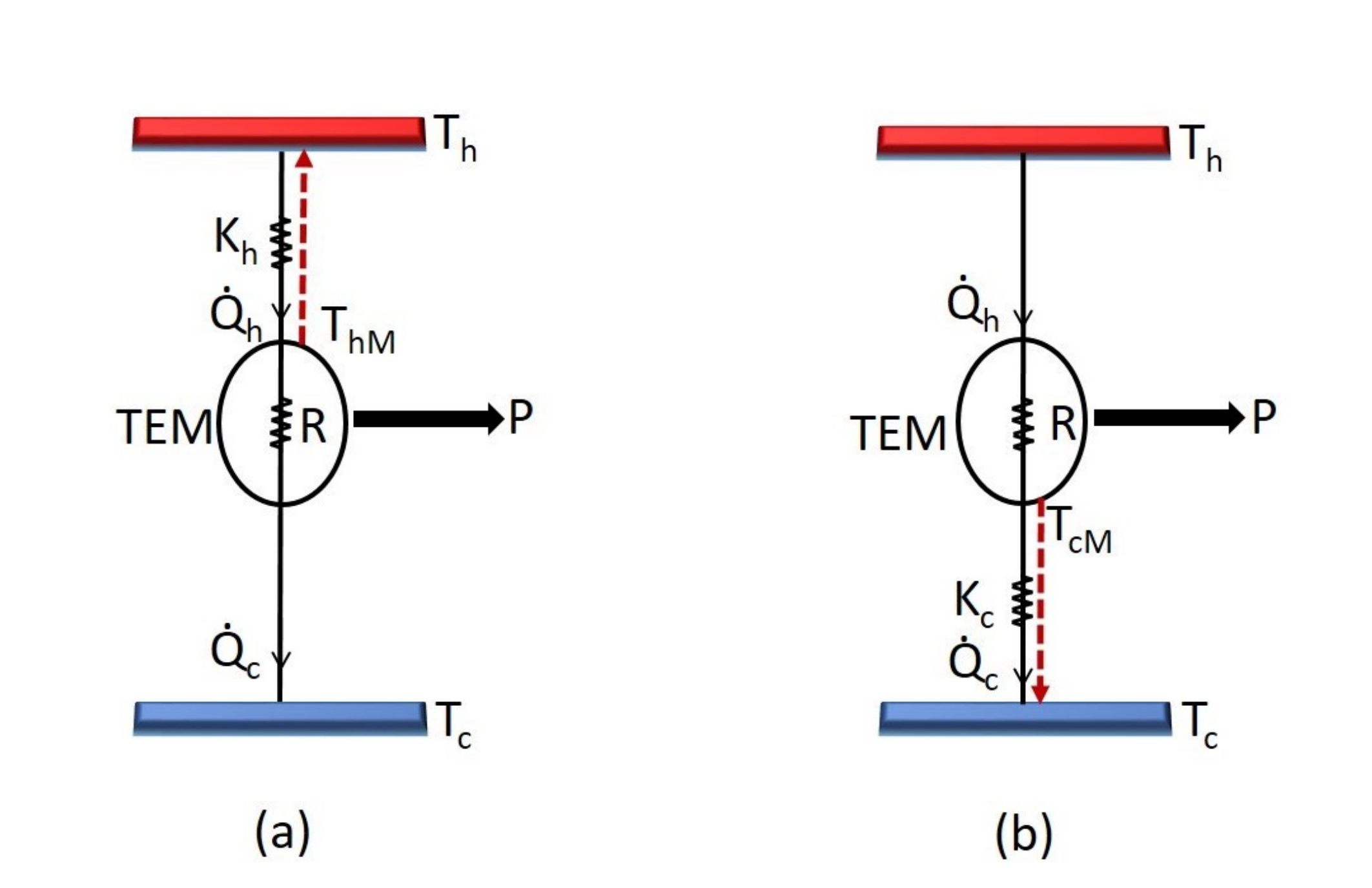}
\caption{ (a) Case 1: The external and internal 
irreversibilities are assumed on the side of hot reservior.
(b) Case 2: Both external and internal
irreversibilities are considered towards the side of cold reservior.}
\label{fig2}
\end{figure}
On the other hand, $K_{\rm ext} \gg K_{\rm int}$ implies $v\to \infty$, 
or $K_{h} \to \infty$ for a finite $R$. (Here also, $R\to \infty$ would imply a
 vanishing power, and we exclude this possibility). Then
 the efficiency reaches its upper bound, 
  ${\eta_{C}^{}}/({2-\eta_{C}^{}})$. Therefore,
   for $0 <v <\infty$, ${\eta }^*$ lies in the range
\be
1-\sqrt{1-\eta_{C}^{}} \le \eta^{*} \le \frac{\eta_{C}^{}}{2-\eta_{C}^{}}.
\label{effrange1}
\ee
In this model, there are no dissipative losses on cold side of TEM. Further, the
dissipation due to Joule heating acts as positive feedback for device 
performance \cite{SISMAN1995,Apertet2013B} and helps to increase the EMP.

Next, the maximum power is given by the expression
\be
P^{*}_{v}=K_{h}^{} T_{h}^{}\left[ 1+2v\theta+\theta-2 \sqrt{\theta(v+1)(v\theta+1)}\right],
\label{pv}
\ee
which has the following limiting behaviour:
\bea 
P_{v \to 0}^{*} &=& K_{h}^{}T_{h}^{}(1-\sqrt{1-\eta_{C}^{}} )^2,
\label{Pat0}\\
P_{v \to \infty}^{*} &=& \frac{\alpha^2 T_{h}^{2}}{R} \frac{\eta_{C}^{2}}{4}.
\label{Patinf}
\eea
\section{Case 2: Finite $K_{c}{}$ with $\omega=1$}
In this case, the external irreversibility is considered only on the cold side.
 Further, we assume $\omega=1$, i.e. Joule heat flows into the cold reservoir 
as shown in Fig. 2b. On the cold side, 
the thermal flux is
\bea
\dot{Q}_{c}^{}=\frac{\alpha T_{c}^{} K_{c}^{}I+R K_{c}^{} I^{2}}{K_{c}^{}-\alpha 
I},
\label{LT3} 
\eea
and 
 \be
 T_{cM}=\frac{K_{c}T_{c}+R I^2}{K_{c}-\alpha I}.
 \ee
Thermal flux on the (reversible) hot side of TEG is		
\be
\dot{Q}_{h}^{}=\alpha T_{h}^{} I. \label{LT4}
\ee
Then, optimizing power w.r.t $I$, gives EMP as 
\begin{equation}
\eta^{*}= v + 1- \sqrt{(v+1)(v+ \theta)}, \label{RE12}
\end{equation}
where  $v = R K_c/ (\alpha^2 T_h)$, with $K_{\rm ext}=K_{c}$ and $K_{\rm int}=
{\alpha^2 T_{h}}/{R}$. Then, it can be shown that for $0 < v <\infty$, the
corresponding  value of $\eta^{*}$ lies in the range:
\begin{equation}
\frac{\eta_{C}^{}}{2} \le \eta^{*} \le 1-\sqrt{1-\eta_{C}^{}}.
\label{effrange2}
\end{equation}
In this model, the dissipation due to Joule heating results in decreasing the
EMP of TEG. 

Further, the maximum output power is 
\be
P^{*}_{v}=K_{c}^{}T_{h}^{}\left[ 1+2v+\theta-2\sqrt{(v+1)(v+\theta)}\right].
\ee
The limiting behaviour of $P^{*}_{v}$ is
\be 
P_{v \to 0}^{*} = K_{c}^{}T_{h}^{}(1-\sqrt{1-\eta_{C}^{}} )^2
\label{pat0}
\ee
and  $P_{v \to \infty}^{*}$ as in Eq. (\ref{Patinf}).
    \section{Case 3: Finite $K_{h}^{}$ with $\omega=1$} \label{section4}
\begin{figure}
    	\centering
   	\includegraphics[width=9cm,height=6.5cm]{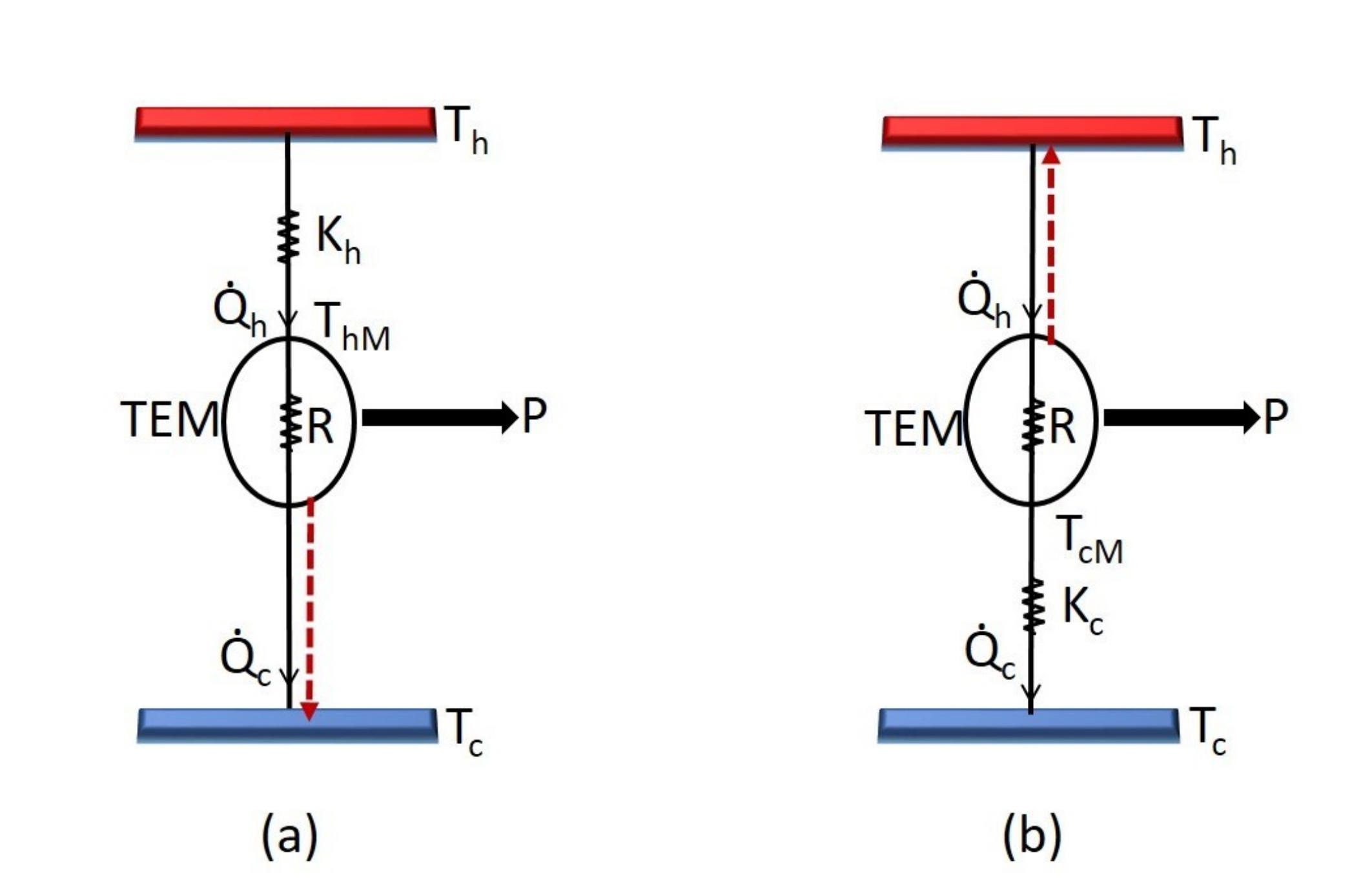}
   	\caption{(a) Case 3: The external irreversibility on the
    		hot side of TEM and Joule heat is dumped on the cold side of TEM.
    		(b) Case 4: The external irreversibility on the cold
	  side of TEM and Joule heat is dumped
	 into the hot reservoir.}
    	\label{fig3}
   \end{figure}

In this case, the system is simplified by considering external irreversibility
on the hot side and the flow of Joule heat to the cold side of TEG, as 
shown in Fig. \ref{fig3}(a). Now, the heat fluxes on the hot and cold sides of 
TEM are given by
    \begin{eqnarray}
    \dot{Q}_{h}^{} &=& \frac{\alpha K_{h}^{} T_{h}^{}I}{K_{h}^{}+\alpha I},   
    \label{LT111} \\
    \dot{Q}_{c}^{} &=& \alpha T_{c}^{} I+R I^2, \label{LT222}
    \end{eqnarray}
    and so the power output can be written as function of $I$. 
    Optimizing power w.r.t $I$, the EMP comes in the form 
    \begin{equation}
    \eta^{*}=1-\theta-(v+\theta)A-v A^2,
    \end{equation}
      where
      	\begin{equation} \label{eq31}
      	      A =  \frac{1}{6v}\left[ x^{1/3}
    +\frac{(2v-\theta)^2}{x^{1/3}}
     	-(4v+\theta) \right]  
           	\end{equation}
       with $x =  (54v^2+(2v-\theta)^3+6\sqrt{3}v\sqrt{27v^2+(2v-\theta)^3})$,
   which has been obtained using Mathematica software.  
    In the above, $v = R K_h /(\alpha^2 T_h)$. Again, it is concluded that
      for $0 < v  < \infty$, the value of $\eta^{*}$ lies in the same range 
      as Eq. (\ref{effrange2}).
   
The maximum output power is 
    \be
    P^{*}_{v}=K_{h}^{} T_{h}^{}\frac{A}{A+1}\left[ 1-\theta-(v+\theta)A-v 
A^2\right],
    \ee
whose limiting behavior is identical with Eqs. (\ref{Pat0}) and (\ref{Patinf}).
\section{Case 4: Finite $K_{c}^{}$ with $\omega=0$} \label{section6}
In this case, the external irreversibility is considered  only on cold side,
 while internal irreversibility due to Joule heating considered to be on hot 
side of TEG \cite{Apertet2014}, as shown in Fig. \ref{fig3}(b). 
Thermal flux on the hot side of TEG is then 
\begin{eqnarray}
\dot{Q}_{h}^{}=\alpha T_{h}^{}I-R I^{2},  \label{LT1111}
\end{eqnarray}
while the thermal flux on the cold side of TEG is given by 
$\dot{Q}_{c}^{}=\alpha T_{cM} I$, where
\be
T_{cM} = \frac{ K_{c}^{} T_{c}^{}}{K_{c}^{}-\alpha I}.
\ee
Therefore, 
\be
\dot{Q}_{c}^{}=\frac{\alpha K_{c}^{} T_{c}^{} I}{K_{c}-\alpha I}.
\label{LT2222}
\ee
Then, the EMP is given by 
\begin{equation}
\eta^{*}=1-\frac{\theta^2}{(vB-1)(B-\theta)},
\label{RE1}
\end{equation}
where
\begin{equation}
    B =\frac{1}{6v}\left[ 4v \theta+1-y^{1/3}-
 \frac{(1-2v\theta)^2}{y^{1/3}}\right],  
\end{equation}
with
\begin{equation*}
y=54v^{2}\theta^3+(2v\theta-1)^3+6\sqrt{3}v\theta
\sqrt{\theta((2v\theta-1)^{3}+27v^{2} \theta^3)}
\end{equation*}
and $v = R K_c/(\alpha^2 T_c)$. 
Then,  for $0 < v < \infty$, $\eta^{*}$ lies in the same range as 
Eq. (\ref{effrange1}).

The maximum output power is 
\be
P^{*}_{v}=K_{c}^{} T_{h}^{}\frac{B(1-vB)}
{\theta} \left[1-\frac{\theta^2}{(vB-1)(B-\theta)}\right].
\ee
For $v \to 0$, the above expression reduces to Eq. (\ref{pat0}) and, for $v \to 
\infty$, it equals Eq. (\ref{Patinf}). 
\section{Case 5: Finite $K_{h}=K_{c}$ with $\omega=1/2$} 
We have so far considered an external  irreversibility which is towards either
hot or cold reservoir. Similarly, the internal irreversibility due to Joule 
heating was assumed to be rectified, thus rejecting itself 
in one of the two reservoirs only. 
However, an interesting special case arises with the simultaneous presence of 
internal as well as external irreversibilities. This is when we treat the external 
irreversibilities at the hot and the cold contact in a symmetric manner, 
as well as consider a symmetric dumping of Joule heat at each reservoir 
\cite{Apertet2012B}. More precisely, it means 
 setting $K_{h} = K_{c} = K_0$, and $\omega =1/2$. In this case, we have 
 \be 
K_{\rm ext} = \frac{K_0}{2}, \quad K_{\rm int} =  
\frac{\alpha^2 (T_{h} + T_c)}{2R},
\ee
and so parameter $v =  R K_0 /[\alpha^2 (T_h + T_c)]$. 
This model is also solvable for optimal power, and yields EMP of 
the form: 
\be
\eta^{*} = \frac{(2-v-v\theta)-\sqrt{(v+v\theta+2)(v+v 
\theta+2\theta)}}{2(1-\theta)-v(3+4\theta+\theta^2)} (1-\theta).
\ee
In the limit, $v \to 0$, it reduces to the endoreversible model, yielding the
$\eta_{\rm ext}$ value. When $v \to \infty$, the internal irreversibility
becomes dominant as compared to external irreversibility,  
and we obtain Eq. (\ref{SS}) with $\omega=1/2$. Thus, for 
$0 < v <\infty$, $\eta^{*}$ lies in the range
\be
1-\sqrt{1-\eta_{C}^{}}\le \eta^{*} \le \frac{2\eta_{C}^{}}{4-\eta_{C}^{}}.
\ee 
The maximum output power is 
\be
P^{*}_{v}=\frac{K_{0}  T_{h}^{}}{2}\left[ 1 + v + \theta + v 
\theta-\sqrt{4\theta+v(2+v)(1+\theta)^2} \right]. 
\ee
Maximum power has the following limiting expressions:
\bea 
P^{*}_{v\to 0} & =& \frac{1}{2} K_0 T_{h}^{}(1-\sqrt{1-\eta_{C}^{}})^2,
\eea
and for $v \to \infty$, it reduces to Eq. (\ref{Patinf}).
\section{Discussion}
We have analyzed the maximum power conditions of a TEG taking 
into account two kinds of irreversibilities, internal due to Joule heating 
and external due to non-ideal thermal contacts with heat exchangers. 
In this treatment, we have assumed tight coupling between the fluxes,
and so neglected any heat leakages. Thus, we have considered 
special instances of these models which turn out to be exactly
solvable. The main focus has been the derivation of compact 
expressions for EMP as well as maximum power.
All expressions for EMP 
obey well-defined upper and lower bounds when 
the ratio of external to internal
 thermal conductances (parameter $v$) vanishes or becomes very large. 
 
\begin{figure}
\centering
\includegraphics[width=8cm]{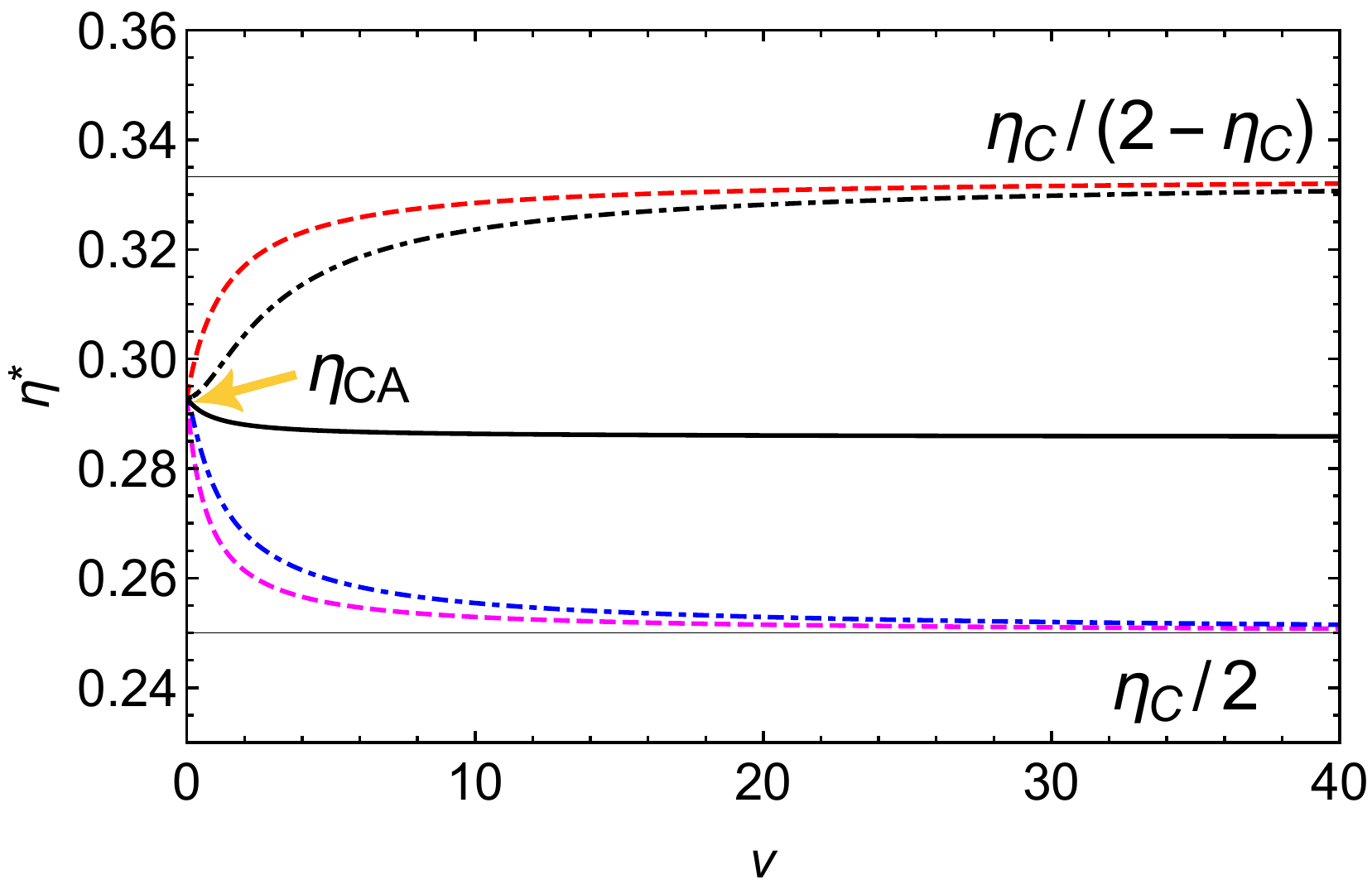}
\caption{EMP as a function of $v$ for various cases, with $\eta_{C}=0.5$.  For 
$v \to 0$, $\eta_{\rm CA}$ is obtained in all the cases. Top and bottom dashed 
lines show case 1 and 2 respectively, while dot-dashed lines show case 4 and 3. 
For large $v$, these cases approach upper or lower bound for EMP. The middle 
solid line shows case 5.}
	\label{fig4}
\end{figure}

 For better comparison, EMPs in all the five cases have been plotted
 versus $v$, in Fig. \ref{fig4}. The vanishing of 
 $v$ represents so-called endoreversible limit 
 \cite{CA1975, Josefsson2018} and its large
 values represents so-called exoreversible limit 
 \cite{SchmiedlSeifert2008, Chen1989, Johal2018}. 
 Thus, we are able to highlight interpolating
 behavior of EMP in a TEG, by studying various special cases. Similarly, we have 
 discussed expressions for optimal power in all the cases. It is observed that 
 the maximum power in all cases, approaches the same limiting 
 value (Eq. (\ref{Patinf})), as $v\to \infty$. 

 Apart from the exact expressions, the 
 universal properties of EMP near equilibrium ($\theta \approx 1$)
 are also of interest. Thus, we find that the 
 EMP, in all the cases, follows the linear response universality 
 \cite{VandenBroeck2005, Iyyappan2017}
  given by $\eta_{C}^{}/2$, and is independent
 of parameter $v$. This is expected from the lower and upper bounds
 of EMP as they also obey the same feature.
 The coefficients of the second order terms, in general,
 depend on $v$. However, in the symmetric case 5, we find that the 
 second order term is also universal and is given by 
$\eta_{C}^{2}/8$ \cite{Esposito2009}. 
\begin{figure}
\centering
\includegraphics[width=8cm]{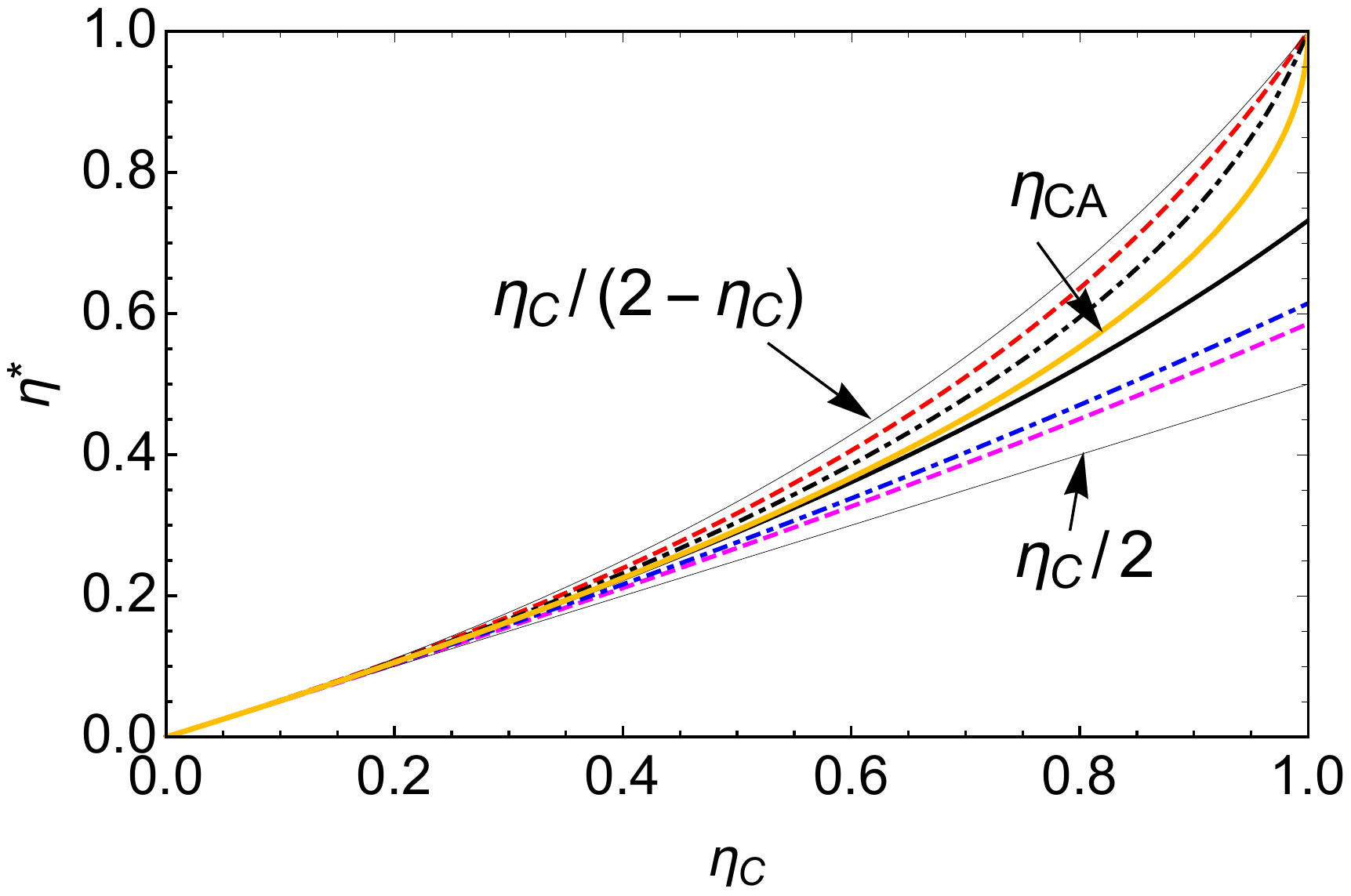}
\caption{Efficiency at maximum power for $v=1$, as a function of 
$\eta_{C}^{}$
for various cases as in Fig. 4. As reference, 
the upper and lower bounds of EMP are marked, 
along with Curzon-Ahlborn (CA) value.}
\label{fig5}
\end{figure}
Further, the second order terms in different
cases show an interesting trend if we set $v=1$, i.e. consider equal magnitudes 
for 
the external and the internal irreversibilities in each case. 
This is also known as the thermal impedence matching condition \cite{Apertet2012A}. 
Note that this implies a different operating point for each case. 
These cases are depicted in Fig. \ref{fig5}.

\begin{table}
 \begin{tabular}{|c|c|c|c|c|}
\hline 
Case & $\omega$ &  $v = \frac{K_{\rm ext}}{K_{\rm int}}$ & EMP at $v=1$
&  $R_{\rm load}/R$ at $v=1$\\
\hline 
1 &  0 &  $\frac{R K_h}{\alpha^2 T_c}$  
&  $\frac{ \eta_{C}^{}}{2}+\frac{6}{32}\eta_{C}^{2}  + O[\eta_{C}^3]$
& $2$ \\
2 &  1 & $\frac{R K_c}{\alpha^2 T_h}$  
&  $\frac{ \eta_{C}^{}}{2}+\frac{2}{32}\eta_{C}^{2}  + O[\eta_{C}^3]$
& $2$ \\
3 &  1 & $\frac{R K_h}{\alpha^2 T_h}$  
&  $\frac{ \eta_{C}^{}}{2}+\frac{3}{32}\eta_{C}^{2}  + O[\eta_{C}^3]$
& $2+\frac{\eta_{C}}{2}+\frac{3}{32}\eta_{C}^{2}$ \\
4 &  0 & $\frac{R K_c}{\alpha^2 T_c}$  
&  $\frac{ \eta_{C}^{}}{2}+\frac{5}{32}\eta_{C}^{2}  + O[\eta_{C}^3]$
& $2-\frac{\eta_{C}^{}}{2}+\frac{19}{32}\eta_{C}^{2}$ \\
5 & $\frac{1}{2}$ & $\frac{R K_0}{\alpha^2 (T_h+T_c)}$  
& $\frac{ \eta_{C}^{}}{2}+\frac{4}{32}\eta_{C}^{2}  + O[\eta_{C}^3]$
& $2-\frac{ 1}{32}\eta_{C}^{2}$ \\
\hline
 \end{tabular}
 \caption{EMP upto second order term for all cases at 
 	 thermal impedence matching condition ($v=1$), along with  
 	 equivalent load resistance corresponding to maximum 
 	 power.}
\end{table}

The quantitative behavior of EMP upto second order terms is further shown in Table I.
Thus, amongst the studied cases, when  external and 
internal thermal conductances are same ($v=1$), the coefficients
of second order terms show a change in steps of $\eta_{C}^2/32$.
More specifically, we may compare cases 1 and 4, for which $\omega=0$.
At thermal impedence matching condition ($v=1$), 
we have $K_{\rm ext} \equiv \alpha^2 T_c /R$
 is equal to $K_h$ or $K_c$, respectively. From  Table I, we note
that upto second order in $\eta_C$, the EMP decreases from case 1 to case 4,
by an amount $\eta_{C}^{2}/32$. Similarly, we may compare cases 3 and 2, for which 
$\omega =1$. At $v=1$, we have $\alpha^2 T_h /R$ 
equal to $K_h$ or $K_c$, respectively. Here also, the 
EMP decreases by $ \eta_{C}^{2}/32$. Thus, given that Joule heat is dumped 
entirely on one side, when the external irreversibility 
is shifted from hot side to cold side, the EMP decreases by  $\eta_{C}^{2}/32$,
upto second order. 

An alternate way of comparing EMP at $v=1$, is to take, say, cases 
1 and 3, for which the external irreversibility is $K_h$.
Then at $v=1$, $K_h$ is equal to $\alpha^2 T_c/R$ or $\alpha^2 T_h /R$, 
respectively. In this situation, the EMP decreases by an amount $3\eta_{C}^{2}/32$
in going from case 1 to case 3.
Similarly, for cases 4 and 2, for which external irreversibility is
on the cold side, we deduce that at $v=1$, the EMP 
decreases by $3\eta_{C}^{2}/32$, in going from case 4 to case 2. 
Thus, with external irreversibility on only one side (finite $K_h$ or $K_c$), 
the EMP decreases by a step of $3\eta_{C}^{2}/32$, when the Joule heating 
is shifted entirely from the hot side to the cold side.  
\subsection{Equivalent electrical circuit of TEG}
Finally, we discuss the operational meaning of optimization of 
power with respect to the electric current. 
There is an external load resistance $R_{\rm load}$ in the TEG circuit
and the current flowing through it can be varied  by tuning this load.  
The electric power output by the TEG can also be given by 
\be
P=I^2 R_{\rm load}. 
\label{LT5}
\ee
Here, we are interested in the value of the load at the  
maximum power output. In particular, for case 1,  
by equating Eq. (\ref{LT5}) with the expression of power 
(using Eq. (\ref{LT1}) and (\ref{LT2})),
we obtain following expression for $R_{\rm load}$
\be
R_{\rm load}= \frac{\alpha K_{h}(T_{h}-T_{c})-
(R K_{h}+\alpha^2 T_{c})I}{I(K_{h}+\alpha I)}. 
\label{LT6}
\ee
In order to find the $R_{\rm load}$ at maximum power in case 1, 
we substitute the following $I^{*}$ at maximum power 
\be 
I^*=\frac{K_{h}}{\alpha}\left[ \sqrt{\frac{1+v \theta}{\theta(1+v)}}-1\right]. 
\ee
Now, plug in the value of $I^{*}$ in Eq. (\ref{LT6}). 
The value of $R_{\rm load}$ at maximum power becomes,
\be
R^*_{\rm load}=R\left[ 1+\frac{1}{v}\right], 
\label{rload1}
\ee
where $v=R K_{h}/(\alpha^2T_{c})$.
Operationally, this should be the value of load resistance of 
the TEG works at the maximum power. In the limit, $v\to \infty$, 
when thermal contacts are perfect,
$R_{\rm load}$ tends to $R$.  On the other hand, 
as $v \to 0$, $R_{\rm load}$ is tends to the value $\alpha^2 T_{c}/K_{h}$.
For case 2 also, we obtain the same value of optimal load as in Eq. (\ref{rload1}).
Similarly, for other cases, the value of $R_{\rm load}$ can be calculated. 
These expressions are complicated for the cases 3 and 4. 
We indicate the values of $R_{\rm load}$
in different cases for the impedence matching condition and 
for small temperature differences, in Table I.

\section{Conclusion} 
We have studied optimal power conditions for a few special configurations
of external and internal irreversibilities in a 
one-dimensional TEG model with constant properties. We find explicit expressions
for EMP which are functions only of the ratio of external to internal 
thermal conductances, apart from the ratio of bath temperatures. 
The bounds on EMP are discussed for each case, whereby the model 
approaches either endoreversible or exoreversible limit.
 We have also 
discussed the symmetric case (case 5) and show interpolating 
behavior of EMP between CA efficiency and Schmiedl-Seifert efficiency 
\cite{Apertet2012B,SchmiedlSeifert2008}. Further, interesting 
trend is shown by EMP near equilibrium for the thermal impedence 
matching condition. This also helps to distinguish the relative 
magnitudes of EMP in the various configurations.
From the studied cases, we conclude that higher values of EMP are obtained
when both internal and external irreversibilities are taken
on the hot side. Similarly, lower values of EMP are obtained if 
these irreversibilities are taken on the cold side. Again, dumping of Joule
heat on the hot side raises EMP as compared to dumping
it on the cold side. Thus our analysis provides an 
insight into improving the performance of a TEG.
 Out of five discussed cases, cases 1 and 4 are
recommended for designing a TEG. If one requires TEG with higher efficiency, 
then one goes for a design based on case 1. If
the requirement is of a device with higher output power,
then one can choose a design as in case 4. 
In this choice, the design of specific thermoelectric materials which 
have an ability to rectify Joule heat, also plays an important role. 
Finally, a parallel analysis can be undertaken for thermoelectric refrigerators. 

\section*{Acknowledgements}
JK acknowledges financial support in the form of Senior Research Fellowship 
from IISER Mohali, and useful discussions with I. Iyyappan and Varinder Singh. 


\end{document}